\documentclass{emulateapj}

\def\acknowledgeIRAMPdBI{This work is based on observations carried out with the
 IRAM Plateau de Bure Interferometer. IRAM is supported by INSU/CNRS (France),
 MPG (Germany) and IGN (Spain).\ }

\def\mbox{\hbox}           

\def\CHp{\ifmmode \mbox{\rm CH}^+           
         \else {\rm CH}$^+$                 
         \fi}

\def\CO#1{\ifnum#1=0                    
           \ifmmode \mbox{\rm CO}
           \else {\rm CO}
           \fi
          \else
           \ifnum#1<15
            \ifmmode ^{#1}\mbox{\rm CO}
            \else $^{#1}${\rm CO}
            \fi
           \else
            \ifmmode \mbox{\rm C}^{#1}\mbox{\rm O}
            \else {\rm C}$^{#1}${\rm O}
            \fi
           \fi
          \fi}

\def\COp{\ifmmode \mbox{\rm CO}^+           
         \else {\rm CO}$^+$                 
         \fi}

\def\CS#1{\ifnum#1=0                    
           \ifmmode \mbox{\rm CS}
           \else {\rm CS}
           \fi
          \else
           \ifnum#1<15
            \ifmmode ^{#1}\mbox{\rm CS}
            \else $^{#1}${\rm CS}
            \fi
           \else
            \ifmmode \mbox{\rm C}^{#1}\mbox{\rm S}
            \else {\rm C}$^{#1}${\rm S}
            \fi
           \fi
          \fi}

\def\HCOp{\ifmmode \mbox{\rm HCO}^+          
          \else {\rm HCO}$^+$                
          \fi}

\def\HCtN{\ifmmode \mbox{\rm HC}_3\mbox{\rm N}     
          \else {\rm HC}$_3${\rm N}                
          \fi}

\def\Hp{\ifmmode \mbox{\rm H}^+           
        \else {\rm H}$^+$                 
        \fi}
\def\HtOp{\ifmmode \mbox{\rm H}_2\mbox{\rm O}^+   
          \else {\rm H}$_2${\rm O}$^+$            
          \fi}
\def\HthOp{\ifmmode \mbox{\rm H}_3\mbox{\rm O}^+   
              \else {\rm H}$_3${\rm O}$^+$         
              \fi}
\def\Hthp{\ifmmode \mbox{\rm H}_3^+         
             \else {\rm H}$_3^+$            
             \fi}

\def\Ht{\ifmmode \mbox{\rm H}_2              
        \else {\rm H}$_2$                    
        \fi}

\def\HtCO{\ifmmode \mbox{\rm H}_2\mbox{\rm CO} 
          \else {\rm H}$_2${\rm CO}            
          \fi}

\def\HtO{\ifmmode \mbox{\rm H}_2\mbox{\rm O} 
         \else {\rm H}$_2${\rm O}            
         \fi}
\def\HteO{\ifmmode \mbox{\rm H}_2^{18}\mbox{\rm O} 
          \else {\rm H}$_2^{18}${\rm O}            
          \fi}

\def\OHp{\ifmmode \mbox{\rm OH}^+   
         \else {\rm OH}$^+$         
         \fi}
\def\ion#1#2{\ifmmode \mbox{{\rm #1}}\,\mbox{{\sc #2}} 
        \else {\rm #1}$\,${\sc #2}
        \fi}

\def\CII{\ion{C}{ii}}

\def\rec#1#2{\if#2a                            
              \ifmmode \mbox{{\rm #1}}\alpha   
              \else {\rm #1}$\alpha$
              \fi
             \fi
             \if#2b
              \ifmmode \mbox{{\rm #1}}\beta
              \else {\rm #1}$\beta$
              \fi
             \fi
             \if#2g
              \ifmmode \mbox{{\rm #1}}\gamma
              \else {\rm #1}$\gamma$
              \fi
             \fi}

\newcommand{\tabref}[1]{Table~\protect\ref{#1}}
\newcommand{\figref}[1]{Fig.~\protect\ref{#1}}

\newcommand{\eqref}[1]{Eq.~$\left(\protect\ref{#1}\right)$}

\def\mbox{\hbox}           

\def\deg{\ifmmode ^\circ                
         \else $^\circ$
         \fi
         \hskip -0.1truecm}
\def\degd#1.#2{                         
               \ifmmode {#1^{\hskip 0.05em\circ}\hskip-0.42em.\hskip0.08em#2}
               \else {#1$^{\hskip 0.05em\circ}\hskip-0.42em.\hskip0.08em$#2}
               \fi
              }
\def\mind#1.#2{                         
               \ifmmode {#1^{\hskip 0.05em\prime}\hskip-0.35em.\hskip0.05em#2}
               \else {#1$^{\hskip 0.05em\prime}\hskip-0.35em.\hskip0.05em$#2}
               \fi
              }
\def\secd#1.#2{                         
               \ifmmode {#1^{\prime\prime}\hskip-0.46em.\hskip0.12em#2}
               \else {#1$^{\prime\prime}\hskip-0.46em.\hskip0.12em$#2}
               \fi
              }
\def\timsecd#1.#2{                      
                  \ifmmode {#1^{\rm s}\hskip-0.39em.\hskip0.08em#2}
                  \else {$#1^{\rm s}\hskip-0.39em.\hskip0.08em#2$}
                  \fi
                 }
\def\hms#1h#2m#3s{                      
                  \relax
                  \ifmmode #1^{\rm h}\,#2^{\rm m}\,#3^{\rm s}
                  \else \hbox{$#1^{\rm h}\,#2^{\rm m}\,#3^{\rm s}$}
                  \fi
                 }
\def\dms#1d#2m#3s{                      
                  \relax
                  \ifmmode #1^\circ\,#2^{\prime}\,#3^{\prime\prime}
                  \else \hbox{$#1^\circ\,#2^{\prime}\,#3^{\prime\prime}$}
                  \fi
                 }
\def\dmsd#1d#2m#3.#4s{                  
                      \relax
                      \ifmmode #1^\circ\,#2^{\prime}\,#3^{\prime\prime}
                               \hskip-0.46em.\hskip0.12em#4
                      \else \hbox{$#1^\circ\,#2^{\prime}\,#3^{\prime\prime}
                            \hskip-0.46em.\hskip0.12em#4$}
                      \fi
                     }
\def\hm#1h#2m{                          
              \relax
              \ifmmode #1^{\rm h}\,#2^{\rm m}
              \else \hbox{$#1^{\rm h}\,#2^{\rm m}$}
              \fi
             }
\def\dm#1d#2m{                          
              \relax
              \ifmmode #1^\circ\,#2^{\prime}
              \else \hbox{$#1^\circ\,#2^{\prime}$}
              \fi
             }
\def\hmsd#1h#2m#3.#4s{                  
                      \relax
                      \ifmmode #1^{\rm h}\,#2^{\rm m}\,#3^{\rm s}
                               \hskip-0.39em.\hskip0.08em#4
                      \else \hbox{$#1^{\rm h}\,#2^{\rm m}\,#3^{\rm s}
                            \hskip-0.39em.\hskip0.08em#4$}
                      \fi
                     }
\def\hmd#1h#2.#3m{                  
                  \relax
                  \ifmmode #1^{\rm h}\,#2^{\rm m}
                           \hskip-0.55em.\hskip0.22em#3
                  \else \hbox{$#1^{\rm h}\,#2^{\rm m}
                        \hskip-0.55em.\hskip0.22em#3$}
                  \fi
                 }
\def\mg{\relax                          
        \ifmmode ^{\rm m}
        \else $^{\rm m}$
        \fi
       }
\def\mgd#1.#2{                          
              \relax
              \ifmmode #1^{\rm m}
                       \hskip-0.55em.\hskip0.22em#2
              \else \hbox{#1$^{\rm m}
                    \hskip-0.55em.\hskip0.22em$#2}
              \fi
             }

\def\la{\mathrel{\hbox{\rlap{\hbox{\lower4pt\hbox{$\sim$}}}\hbox{$<$}}}}
\def\ga{\mathrel{\hbox{\rlap{\hbox{\lower4pt\hbox{$\sim$}}}\hbox{$>$}}}}

\def\unitspace{\;}                      

\def\un#1{\ifmmode \unitspace\mbox{\rm #1} 
          \else $\unitspace$#1
          \fi}
\def\pun#1#2{\ifmmode \unitspace\mbox{\rm #1}^{#2} 
             \else $\unitspace$#1$^{#2}$
             \fi}
\def\per#1{\ifmmode \unitspace\mbox{\rm #1}^{-1} 
           \else $\unitspace$#1$^{-1}$
           \fi}

\def\GHz{\un{GHz}}                    
\def\Jy{\un{Jy}}                      
\def\K{\un{K}}                        
\def\kms{\un{km}\pun{s}{-1}}          
\def\Lsun{\ifmmode \un{L}_{\odot}     
          \else $\un{L}_{\odot}$
          \fi}
\def\MHz{\un{MHz}}                    
\def\Msun{\ifmmode \un{M}_{\odot}     
          \else $\un{M}_{\odot}$
          \fi}
\def\muJy{\ifmmode \unitspace\mu\mbox{\rm Jy} 
          \else $\unitspace\mu$Jy
          \fi}
\def\mum{\ifmmode \unitspace\mu\mbox{\rm m} 
         \else $\unitspace\mu$m
         \fi}
\def\pc{\un{pc}}                      
\def\pcmcub{\pun{cm}{-3}}             
\def\pcmsqu{\pun{cm}{-2}}             
\def\ps{\pun{s}{-1}}                  
\def\sqarcsec{\ifmmode \unitspace\Box''    
              \else $\unitspace\Box''$     
              \fi} 

\def\Bp{\relax                            
        \ifmmode B_{||}                   
        \else $B_{||}$
        \fi}
\def\Bt{\relax                            
        \ifmmode B\!_{\perp}              
        \else $B\!_{\perp}$               
        \fi}
\def\Dec{\relax                           
          \ifmmode \mbox{\rm Dec}
          \else Dec
         \fi}
\def\Eup{\relax                           
         \ifmmode E_{\rm u}               
         \else $E_{\rm u}$
         \fi}
\def\Gcr{\relax                           
         \ifmmode \Gamma\!_{\rm cr}       
         \else $\Gamma\!_{\rm cr}$
         \fi}
\def\ICII{\relax                          
          \ifmmode I_{[\CII]}             
          \else $I_{[\CII]}$
          \fi}
\def\LHtwo{\relax                                 
           \ifmmode L_{\mbox{\rm\scriptsize H}_2} 
           \else $L_{\mbox{\rm\scriptsize H}_2}$  
           \fi}
\def\LFIR{\relax                           
          \ifmmode L_{\rm FIR}             
          \else $L_{\rm FIR}$
          \fi}
\def\LIR{\relax                           
         \ifmmode L_{\rm IR}              
         \else $L_{\rm IR}$
         \fi}
\def\LLya{\relax                          
          \ifmmode L_{{\rm Ly}\,\alpha}   
          \else $L_{{\rm Ly}\,\alpha}$
          \fi}
\def\mAB{\relax                           
         \ifmmode m_{\rm AB}              
         \else $m_{\rm AB}$
         \fi}
\def\MHtwo{\relax                                 
           \ifmmode M_{\mbox{\rm\scriptsize H}_2} 
           \else $M_{\mbox{\rm\scriptsize H}_2}$  
           \fi}
\def\MHtwodot{\relax                                       
              \ifmmode \dot{M}_{\mbox{\rm\scriptsize H}_2} 
              \else $\dot{M}_{\mbox{\rm\scriptsize H}_2}$  
              \fi}                                         
\def\Mstardot{\relax                      
              \ifmmode \dot{M}_{\ast}     
              \else $\dot{M}_{\ast}$      
              \fi}
\def\nH{\relax                                       
        \ifmmode n_{\mbox{\scriptsize\rm H}}  
        \else $n_{\mbox{\scriptsize\rm H}}$
        \fi}
\def\nHI{\relax                                      
         \ifmmode n_{\mbox{\scriptsize\rm H\,\sc I}} 
         \else $n_{\mbox{\scriptsize\rm H\,\sc I}}$
         \fi}
\def\nHt{\relax                                
         \ifmmode n_{{\mbox{\scriptsize H}}_2} 
         \else $n_{{\mbox{\scriptsize H}}_2}$  
         \fi}
\def\RA{\relax                           
         \ifmmode \mbox{\rm RA}
         \else RA
        \fi}
\def\rhostardot{\relax                         
                \ifmmode \dot{\rho}_{\ast}     
                \else $\dot{\rho}_{\ast}$      
                \fi}
\def\rhoZdot{\relax                          
             \ifmmode \dot{\rho}_{\rm Z}     
             \else $\dot{\rho}_{\rm Z}$      
             \fi}
\def\Tb{\relax                           
        \ifmmode T_{\rm b}               
        \else $T_{\rm b}$
        \fi}
\def\Td{\relax                           
        \ifmmode T_{\rm d}               
        \else $T_{\rm d}$
        \fi}
\def\Te{\relax                           
        \ifmmode T_{\rm e}               
        \else $T_{\rm e}$
        \fi}
\def\Tg{\relax                           
        \ifmmode T_{\rm g}               
        \else $T_{\rm g}$
        \fi}
\def\vhel{\relax                  %
          \ifmmode \qu{v}{hel}    
          \else $\qu{v}{hel}$     
          \fi}
\def\vLSR{\relax                  %
          \ifmmode \qu{v}{LSR}    
          \else $\qu{v}{LSR}$     
          \fi}

\def\sou#1#2{\relax                       
             \ifmmode {\rm #1}\,{\rm #2}  
             \else #1$\,$#2
             \fi}

\def\APM#1{\sou{APM}{#1}}                

\def\Mrk#1{\sou{Mrk}{#1}}                

\def\qu#1#2{\relax                          
            \ifmmode #1_{\rm #2}            
            \else $#1_{\rm #2}$
            \fi}

\slugcomment{in preparation for the Astrophysical Journal Letters}

\shorttitle{Water vapor in APM\,08279+5255}
\shortauthors{Van der Werf et al.}

\begin{document}

\title{Water vapor emission reveals a highly obscured, star forming
nuclear region in the QSO host galaxy APM\,08279+5255 at z=3.9}

\author{Paul P.~van der Werf\altaffilmark{1,2}, A.~Berciano
  Alba\altaffilmark{3,1}, M.~Spaans\altaffilmark{4},
  A.F.~Loenen\altaffilmark{1}, R.~Meijerink\altaffilmark{1},
  D.A.~Riechers\altaffilmark{5}, P.~Cox\altaffilmark{6},
  A.~Wei{\ss}\altaffilmark{7}, F.~Walter\altaffilmark{8}}

\altaffiltext{1}{Leiden Observatory, Leiden University, P.O.~Box 9513, NL-2300
  RA Leiden, The Netherlands} 
\altaffiltext{2}{SUPA, Institute of Astronomy,
  University of Edinburgh, Royal Observatory, Blackford Hill, Edinburgh EH9 3HJ,
  United Kingdom} 
\altaffiltext{3}{ASTRON, P.O.~Box 2, NL-7990 AA Dwingeloo, The
  Netherlands} 
\altaffiltext{4}{Kapteyn Astronomical Institute, University of
  Groningen, P.O.~Box 800, NL-9700 AV Groningen, The Netherlands}
\altaffiltext{5}{Astronomy Department, California Institute of Technology, MC
  249-17, 1200 East California Boulevard, Pasadena, CA 91125, USA}
\altaffiltext{6}{IRAM, 300 Rue de la Piscine, 38406 St.~Martin d'Heres,
  Grenoble, France}
\altaffiltext{7}{Max-Planck-Institut f\"ur Radioastronomie, Auf dem H\"ugel 16,
  Bonn, D-53121, Germany}
\altaffiltext{8}{Max-Planck-Institut f\"ur Astronomie,
  K\"onigstuhl 17, Heidelberg, D-69117, Germany}

\begin{abstract}
  We present the detection of four rotational emission lines of water vapor,
  from energy levels $\Eup/k=101-454\K$, in the gravitationally lensed $z=3.9$
  QSO host galaxy APM\,08279+5255.  While the lowest $\HtO$ lines are
  collisionally excited in clumps of warm, dense gas (density of hydrogen nuclei
  $\nH=(3.1\pm1.2)\times10^6\pcmcub$, gas temperature $\Tg\sim105\pm21\K$), we
  find that the excitation of the higher lines is dominated by the intense local
  infrared radiation field. Since only collisionally excited emission
  contributes to gas cooling, we conclude that $\HtO$ is not a significant
  coolant of the warm molecular gas. Our excitation model requires the
  radiatively excited gas to be located in an extended region of high $100\mum$
  opacity ($\tau_{100}=0.9\pm0.2$).  Locally, such extended infrared-opaque
  regions are found only in the nuclei of ultraluminous infrared galaxies. We
  propose a model where the infrared-opaque circumnuclear cloud, which is
  penetrated by the X-ray radiation field of the QSO nucleus, contains clumps of
  massive star formation where the $\HtO$ emission originates.  The radiation
  pressure from the intense local infrared radiation field exceeds the thermal
  gas pressure by about an order of magnitude, suggesting close to
  Eddington-limited star formation in these clumps.
\end{abstract}


\section{Introduction}

Water is expected to be one of the most abundant molecules in molecular clouds
in galaxies. In cold molecular clouds water is in the form of icy mantles on
dust grains, with total $\HtO$ abundances up to $10^{-4}$ with respect to
hydrogen nuclei \citep{Tielens.etal1991}, thus containing up to 30\% of the
available oxygen atoms. In warm molecular clouds, such as in star forming
galaxies or galaxies with a luminous active galactic nucleus, water can
evaporate from the dust grains when the grain temperature becomes sufficiently
high. $\HtO$ molecules can also be released into the gas phase by
photodesorption in regions exposed to ultraviolet (UV) radiation and by
desorption induced by cosmic rays or X-rays in more obscured regions
\citep{Hollenbach.etal2009}, or by sputtering of grains in shocks. In warm
molecular gas, $\HtO$ can also be formed in the gas phase, through ion-neutral
chemistry in regions with a sufficiently high fractional ionization, or through
neutral-neutral chemistry in regions sufficiently warm that the relevant
activation energy barriers can be overcome. Gas-phase $\HtO$ may play an
important role as a cooling agent of warm, dense molecular clouds
\citep{Neufeld.Kaufman1993,Neufeld.etal1995}. The large Einstein $A$-values of
$\HtO$ rotational transitions lead to high critical densities
($\ga10^8\pcmcub$) so that collisional excitation will only be
effective in very dense gas.

Due to the wet Earth atmosphere, bulk gas-phase water can only be detected from
space, or from distant objects where the cosmological redshift moves the $\HtO$
lines into transparent atmospheric windows.  Despite the detection of a $22\GHz$
water maser in a $z=2.6$ QSO \citep{Impellizzeri.etal2008}, previous searches
for non-maser rotational emission lines of $\HtO$ from high-$z$ objects remained
unsuccessful \citep{Wagg.etal2006,Riechers.etal2006,Riechers.etal2009a}, until
the recent detection of the $2_{0,2}-1_{1,1}$ $\HtO$ emission line from a
gravitationally lensed galaxy at $z=2.3$ \citep{Omont.etal2011}.  However,
neither the molecular excitation mechanism nor the water abundance could be
derived on the basis of the detection of a single optically thick line.

The Herschel Space Observatory has recently enabled the first detections of
$\HtO$ emission lines from two nearby galaxies, with very different results
between the two objects
\citep{VanDerWerf.etal2010,GonzalezAlfonso.etal2010,Weiss.etal2010}.  Spectra of
the lowest $\HtO$ transitions in the nearby starburst galaxy M82 revealed faint
lines with complex spectral shapes, with one of the lines (the $1_{1,1}-0_{0,0}$
para-$\HtO$ ground-state line) in absorption \citep{Weiss.etal2010}. No $\HtO$
emission from higher rotational levels was found \citep{Panuzzo.etal2010}.  In
marked contrast, observations of the nearby Ultraluminous Infrared Galaxy
(ULIRG) and QSO Mrk\,231 revealed a rich spectrum of $\HtO$ lines from upper
level energies $\Eup$ up to $\Eup/k=640\K$ (where $k$ is the Boltzmann
constant), whereas the lower lines connecting to the ground state remained
undetected \citep{VanDerWerf.etal2010}.

In order to search for $\HtO$ in a high-$z$ galaxy in a systematic way, we
undertook a search for four lines of $\HtO$ in the gas-rich $z=3.9$ QSO host
galaxy $\APM{08279+5255}$ \citep{Irwin.etal1998,Lewis.etal1998}.  This object is
gravitationally lensed with a magnification factor $\mu=4$ according to the
model of \citet{Riechers.etal2009a}.  With CO lines detected from rotational
levels up to $J=11$ \citep{Weiss.etal2007,Riechers.etal2009b} and HCN, HNC and
HCO$^+$ lines from levels up to $J=6$
\citep{Wagg.etal2005,GarciaBurillo.etal2006,Riechers.etal2010}, its molecular
medium has been characterized very well.  Analysis of the CO rotational ladder
revealed unusually high excitation \citep{Weiss.etal2007}.  However, a previous
search for the ortho ground-state $1_{1,0}-1_{0,1}$ $\HtO$ line (upper level
energy $\Eup/k=61\K$) was unsuccessful \citep{Wagg.etal2006}.  We therefore
targeted lines of higher excitation, with $\Eup/k$ from $101$ to $454\K$.

\begin{deluxetable*}{l c c c c c l}
\tablewidth{0pt}
\tablecaption{Parameters of H$_2$O line emission from APM\,08279+5255}
\tablehead{
\colhead{} & \colhead{$2_{0,2}-1_{1,1}$} & \colhead{$2_{1,1}-2_{0,2}$} &
\colhead{$3_{2,1}-3_{1,2}$} & \colhead{$4_{2,2}-4_{1,3}$} &
\colhead{$1_{1,0}-1_{0,1}$} & \colhead{}}
\startdata
$\nu_0$\tablenotemark{a} & 987.926 & 752.033 & 1162.911 & 1207.638 & 556.936 &
GHz \\
$\nu_{\rm obs}$\tablenotemark{b} & 201.166 & 153.132 & 236.797 & 245.905 & &
GHz \\
$\Eup/k$\tablenotemark{c} & 101 & 137 & 305 & 454 & 61 & K \\
Flux density calibrator & 3C273 & 3C84 & 3C273 & 0923+392 & & \\
\quad and flux density\tablenotemark{d} & 8.0 & 8.5 & 7.3 & 4.3 & & Jy \\
Synthesized beam\tablenotemark{e} & $\secd 0.58\times\secd0.60$, $33\deg$ & 
$\secd 2.98\times\secd 1.87$, $47\deg$ &
$\secd 0.62\times\secd 0.56$, $48\deg$ & 
$\secd 0.61\times\secd 0.44$, $49\deg$ & & \\
$\sigma$\tablenotemark{f} & 2.5 & 1.7 & 7.1 & 4.3 & & mJy \\
$S_{\rm cont}$\tablenotemark{g} & $16.5\pm0.8$ & $5.4\pm0.3$ & $26.6\pm1.3$ & $31.4\pm2.0$ & & mJy \\
$I_{\rm line}$\tablenotemark{h} & $9.1\pm0.9$ & $4.3\pm0.6$ & $8.0\pm0.9$ &
$7.5\pm 2.1$ & $<0.7$\tablenotemark{n} & Jy$\kms$ \\
$I_{\rm model}$\tablenotemark{i} & 8.9 & 4.5 & 8.4 & 6.5 & 0.25 & Jy$\kms$ \\
$v_0$\tablenotemark{j} & $72\pm 35$ & $47\pm 25$ & $93\pm 16$ &
$50\pm 43$ & & km$\ps$ \\
FWHM\tablenotemark{k} & $492\pm84$ & $480\pm58$ & $312\pm37$ &
$429\pm 96$ & & km$\ps$ \\
$L'_{\rm line}$\tablenotemark{l,m} & $(7.9\pm0.8)\times10^{10}$ &
 $(6.5\pm1.0)\times10^{10}$ & $(5.1\pm0.5)\times10^{10}$ &
 $(4.5\pm1.3)\times10^{10}$ & $<1.9\times10^{10}$ & K$\kms\pc^2$ \\
$L_{\rm line}$\tablenotemark{l,m} & $(2.4\pm0.5)\times10^9$ &
 $(8.9\pm1.3)\times10^8$ & $(2.6\pm0.3)\times10^9$ &
 $(2.6\pm0.7)\times10^9$ & $<1.1\times10^8$ & $L_\odot$ \\[2pt]
\enddata
\tablenotetext{a}{Rest frequency of the line}
\tablenotetext{b}{Observing frequency}
\tablenotetext{c}{Upper level energy of the transition}
\tablenotetext{d}{Assumed flux density of flux calibrator at observing frequency}
\tablenotetext{e}{Full width at half maximum and position angle}
\tablenotetext{f}{R.m.s.\ noise in the integrated spectrum at $\Delta\nu=40\MHz$}
\tablenotetext{g}{Observed integrated continuum flux density}
\tablenotetext{h}{Observed integrated line flux}
\tablenotetext{i}{Modeled integrated line flux}
\tablenotetext{j}{Center velocity of fitted Gaussian relative to $z=3.911$}
\tablenotetext{k}{Full width at half maximum of fitted Gaussian}
\tablenotetext{l}{Using $z=3.911$, $H_0=70\kms\per{Mpc}$, $\Omega_\Lambda=0.73$,
$\Omega_{\rm tot}=1$}
\tablenotetext{m}{Not corrected for lensing}
\tablenotetext{n}{Upper limit from \citet{Wagg.etal2006}}
\label{tab.lines}
\end{deluxetable*}

\notetoeditor{Please set Table 1 upright (not rotated), 2-column width}

\begin{figure}
\plotone{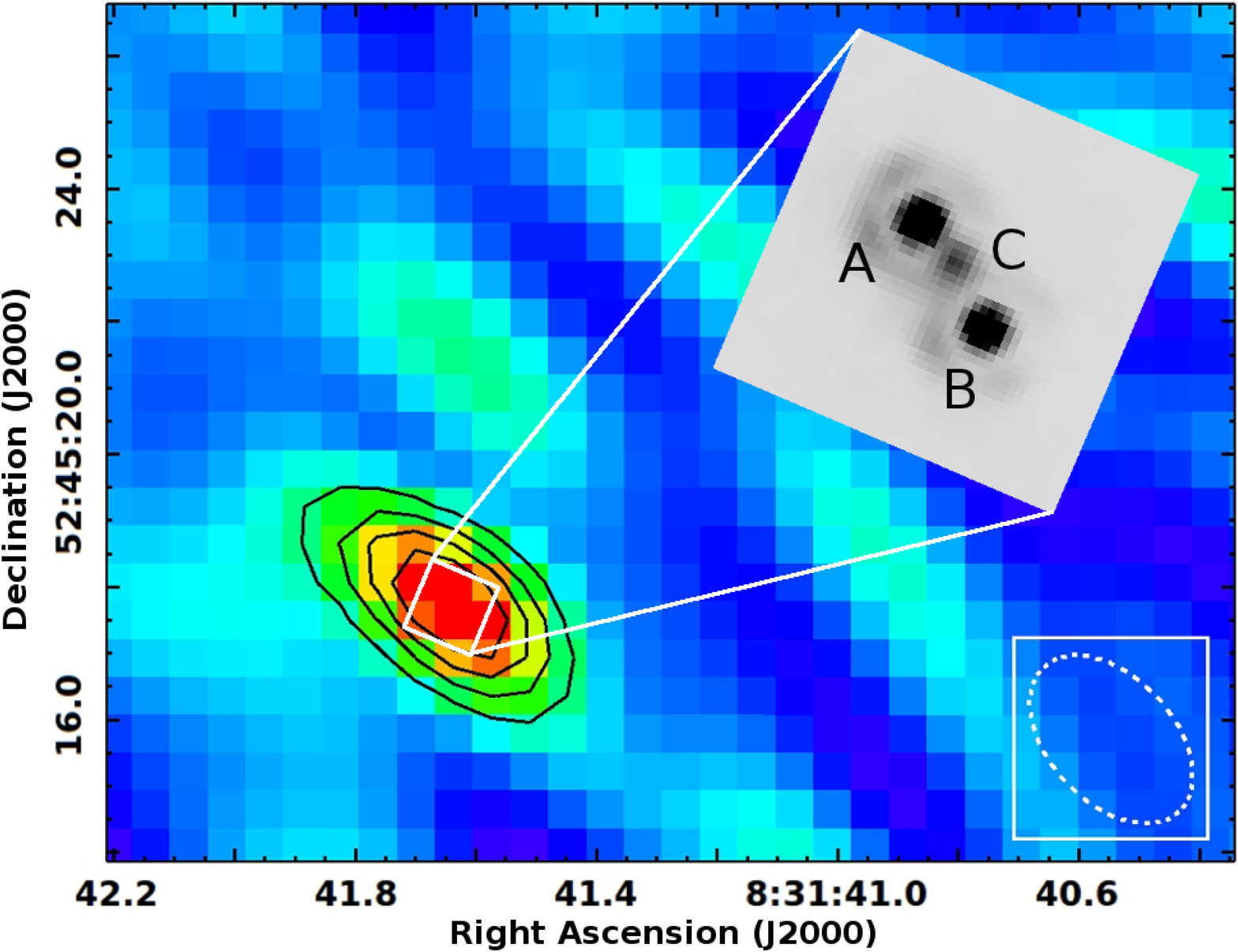}
\caption{H$_2$O emission from APM\,08279+5255. The contoured
  pseudocolour map shows the distribution of the velocity-integrated
  continuum-subtracted $\HtO$ $2_{1,1}-2_{0,2}$ line flux, with contour levels
  of 3, 5, 7 and $9\sigma$, where $\sigma=0.5\Jy\kms\per{beam}$. The dashed
  white ellipse indicates the synthesized beam. The inset presents the NICMOS
  F110W image \citep{Ibata.etal1999} and shows that the gravitationally lensed
  images (the brightest of which are $\secd 0.378$ apart) are fully covered by
  the synthesized beam.
\label{fig.map}}
\end{figure}

\begin{figure}
\plotone{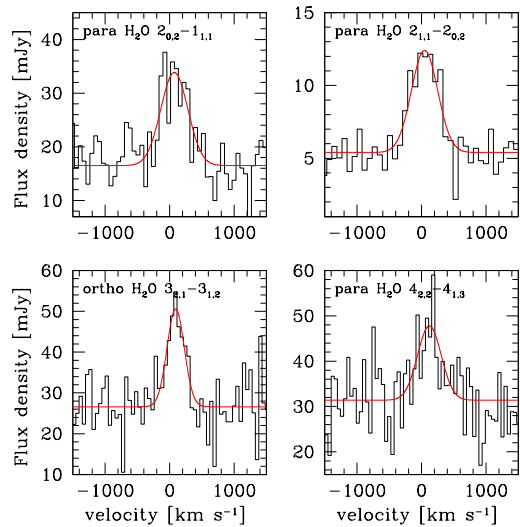}
\caption{Spectra of H$_2$O emission lines from
  $\APM{08279{+}5255}$ at a common frequency resolution of $40\MHz$.
  The horizontal axis shows velocity relative to
  redshift $z=3.911$.  The red curves indicate Gaussian fits to each spectrum.
\label{fig.spectra}}
\end{figure}

\section{Observations and results}
 
We used the IRAM Plateau de Bure Interferometer \citep{Guilloteau.etal1992} with
six antennas in December 2010 and February 2011 to observe the four $\HtO$ lines
listed in \tabref{tab.lines} (which also lists all relevant observational
parameters) in $\APM{08279{+}5255}$.  Observing times (including overheads)
varied from 2 to 5.2\,hours per line. The WIDEX backend was used, providing
$3.6\GHz$ instantaneous bandwidth in dual polarization.  Data reduction using
the GILDAS package included the standard steps of data flagging, amplitude,
phase and bandpass calibration, and conversion into datacubes.  The flux density
scale is accurate to within 15\%. The datacubes were deconvolved using a
CLEAN algorithm \citep{Clark1980}.

All four lines were detected and an image of the flux distribution of the
$2_{1,1}-2_{0,2}$ line is shown in \figref{fig.map}.  Fits to the continuum
$uv$-data were used to calculate the continuum fluxes (reported in
\tabref{tab.lines}) and source sizes. In the high angular resolution
observations the emission was found to be slightly spatially resolved, with
source sizes of approximately $\secd 0.5$, in agreement with high resolution
observations of the $1\un{mm}$ continuum \citep{Krips.etal2007}, and CO $1{-}0$
\citep{Riechers.etal2009a}. Integrated spectra of the four $\HtO$ lines are
shown in \figref{fig.spectra} and parameters of the detected lines are presented
in \tabref{tab.lines}. The most sensitive detection is that of the
$2_{1,1}-2_{0,2}$ line, which displays a symmetric line profile with a full
width at half maximum (FWHM) of $480\pm58\kms$, excellently matching the FWHM
values of $445-520\kms$ of the CO lines \citep{Weiss.etal2007}.  Recently, a low
spectral resolution $1\un{mm}$ spectrum of $\APM{081279{+}5255}$ obtained with
Z-Spec has been released by \citet{Bradford.etal2011}. In the Z-Spec spectrum,
the $3_{2,1}-3_{1,2}$ line is the only $\HtO$ line detected at the $3\sigma$
level, but our flux meaurement for this line is a factor 2 lower. This
discrepancy is similar to that between the CO line fluxes measured by Z-Spec and
by the IRAM $30\un{m}$ telescope, several of which have been independently
confirmed by the IRAM Plateau de Bure Interferometer \citep{Weiss.etal2007}. A
similar discrepancy is found between our flux measurement of the
$4_{2,2}-4_{1,3}$ line and that by \citet{Bradford.etal2011}, although the
latter had a significance of only $2.5\sigma$.  The continuum flux densities
match very well between the two data sets.  Finally, we note that the
serendipitous detection of another $\HtO$ line, the $2_{2,0}-2_{1,1}$ line,
using the IRAM Plateau de Bure Interferometer, was recently reported by
\citet{Lis.etal2011}.

\begin{figure}
\plotone{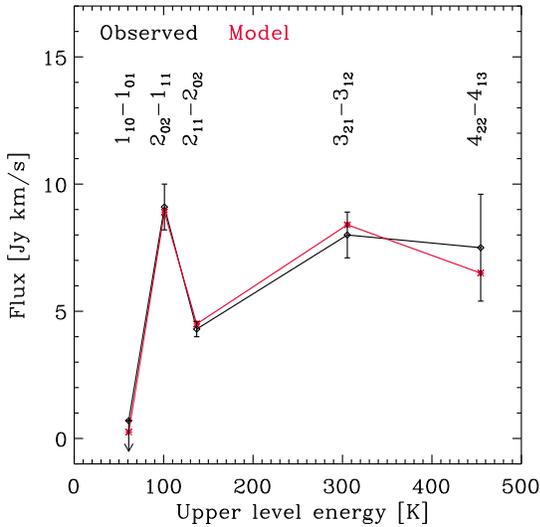}
\caption{Flux in H$_2$O emission lines from $\APM{08279{+}5255}$ as a
  function of upper level energy $\Eup$. Black symbols and error bars
  indicate the observed line fluxes and the previously published upper limit on
  $1_{1,0}-1_{0,1}$ (indicated by a downward arrow). Red crosses
  indicate values produced by our model as described in the text.
\label{fig.emission}}
\end{figure}

\begin{figure}
\plotone{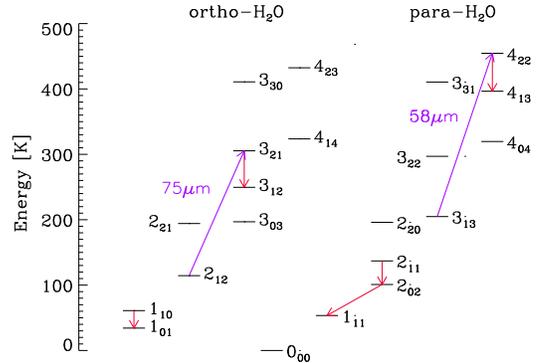}
\caption{Partial H$_2$O energy level diagram. The ortho and
  para-$\HtO$ rotational ladders are indicated, and the transitions
  detected here, as well as the upper limit on the $1_{1,0}-1_{0,1}$
  ortho-$\HtO$ ground state transition are indicated by red
  arrows. Purple arrows indicate the radiative pumping transitions at
  $58$ and $75\mum$ that account for the efficient population of the
  $4_{2,2}$ and $3_{2,1}$ levels.
\label{fig.levels}}
\end{figure}

\section{Far-IR pumped H$_2$O emission}

The distribution of the detected line flux as a function of the energy of the
upper level of the transition is shown in \figref{fig.emission}. The flux
distribution shows considerable emission out to the $4_{2,2}-4_{1,3}$
line, implying that for purely collisional excitation the kinetic temperature
must be of the order of at least $400\K$. Critical densities are approximately
$10^8\pcmcub$ for the lowest lines ($2_{0,2}-1_{1,1}$ and $2_{1,1}-2_{0,2}$) and
higher than $10^8\pcmcub$ for the higher lines. Therefore, any purely
collisionally excited model that can account for the observed $3_{2,1}-3_{1,2}$
and $4_{2,2}-4_{1,3}$ fluxes would produce much stronger emission in the lower
lines, including the $1_{1,0}-1_{0,1}$ level, for which a sensitive upper limit
exists \citep{Wagg.etal2006}. We therefore rule out purely collisional
excitation and consider in addition radiative excitation by an intense
far-infrared (far-IR) radiation field from warm dust, with a temperature
$\Td=220\K$, as derived from the continuum spectral energy distribution
\citep{Weiss.etal2007,Riechers.etal2009a}.

We model these coupled processes using a radiative transfer code
\citep{Poelman.Spaans2005,Poelman.Spaans2006} which computes the statistical
equilibrium populations of all relevant $\HtO$ levels in the ground and first
vibrationally excited state.  Line radiation is transfered in a non-local and
three-dimensional manner through a multi-zone escape probability calculation.
The model consists of dense clumps within a spherical region with a radius of
$100\pc$.  Due to the high critical densities of the $\HtO$ lines, only the
clumps contribute to the $\HtO$ emission, and we have run a model grid where we
varied gas temperatures over the range $\Tg=50-200\K$, and $\HtO$ column
densities over $10^{16}-10^{18}\pcmsqu$.  Clump densities were varied only over
the range $\nH=10^5-10^7\pcmcub$, since higher densities ($\sim10^8\pcmcub$)
would produce low rotational lines ($2_{1,1}-2_{0,2}$ and $2_{0,2}-1_{1,1}$)
much stronger than $3_{2,1}-3_{1,2}$, contrary to what is observed. The
$\HtO$ ortho-to-para ratio was obtained from thermal equilibrium
\citep{Poelman.Spaans2005} and found to be very close to the statistical
equilibrium value of 3.  We fixed the gas-to-dust mass ratio at 100, and the
local turbulent velocity dispersion at $4\kms$, with a total velocity difference
(gradient times length) across the model region of $140\kms$.

In our best fit model, the clumps have a density of hydrogen nuclei
$\nH=(3.1\pm1.2)\times 10^6\pcmcub$, a gas temperature $\Tg=105\pm21\K$, and the
average density of hydrogen nuclei over the $100\pc$ radius model region is
$\langle\nH\rangle=4000\pm1000\pcmcub$, yielding a $100\mum$ continuum optical
depth $\tau_{100}\approx0.9\pm0.2$. The line fluxes predicted by the best fit
model are listed in \tabref{tab.lines} and shown in \figref{fig.emission}.  The
lowest lines (up to $2_{1,1}-2_{0,2}$) are excited mostly by collisions.  The
clump density and temperature derived in our model are determined by these
lines, and the required density is somewhat higher than that derived from the CO
lines \citep{Weiss.etal2007}, which is not surprising given the higher critical
densities of the $\HtO$ lines. In contrast, the $4_{2,2}$ and $3_{2,1}$ levels
are populated exclusively by the absorption of far-IR photons.  As shown in
\figref{fig.levels}, the pumping occurs at $58$ and $75\mum$ rest frame
wavelengths. The intensity of the radiatively excited lines provides a measure
of the intensity of the local far-IR radiation field and therefore of the
optical depth in the spectral region of the pumping wavelengths, and this
determines the $100\mum$ optical depth in our model.  The relative strengths of
the $4_{2,2}-4_{1,3}$ and $3_{2,1}-3_{1,2}$ lines in principle provides
information on the colour temperature of the far-IR radiation field, but for
$\Td>100\K$ the pumping wavelengths are on the Rayleigh-Jeans tail of the warm
dust continuum, making the line ratio insensitive to the dust temperature. Since
only collisionally excited lines contribute to the cooling (i.e., removal of
kinetic energy) of warm molecular gas, we find that cooling by the $\HtO$ lines
is unimportant compared to the cooling by CO rotational lines, in contrast to
the conclusion by \citet{Bradford.etal2011} and to earlier theoretical
suggestions \citep{Neufeld.Kaufman1993,Neufeld.etal1995}. We note that radiative
excitation has also been suggested to drive the intensity of HCN lines in
$\APM{08279{+}5255}$ \citep{GarciaBurillo.etal2006,Weiss.etal2007}. Furthermore,
the upper levels of the [$\CII$] $158\mum$ line and the [$\ion{Si}{ii}$]
$35\mum$ line are also affected by radiative pumping, as shown by rest frame UV
absorption line measurements by \citet{Srianand.Petitjean2000}, who conclude
that the absorbing clouds must be located closer to the extended ($>200\pc$)
source of far-IR radiation than to the UV source.

In our best fit model, the $4_{2,2}-4_{1,3}$ line is only moderately optically
thick (optical depth approximately 1.7), and is thus sensitive to the total
column density of warm ($\Tg>100\K$) water vapor in our model region (unlike the
other lines, which have line center optical depths from 30 to 300). We find a
total (beam-averaged) warm $\HtO$ column density of $2.6\times 10^{17}\pcmsqu$,
i.e., a warm $\HtO$/$\Ht$ adundance of approximately $6\times10^{-7}$,
consistent with UV or X-ray irradiated chemical
models \citep{Meijerink.Spaans2005}.

\section{Implications for the nuclear region of APM\,08279+5255}

The $\HtO$ line ratios observed in $\APM{08279{+}5255}$ are very different from
those observed in prominent star forming regions in the Milky Way. Water lines
from UV irradiated gas (photon-dominated regions or PDRs) in the Milky Way show
thermal level populations \citep{White.etal2010,Habart.etal2010}, dominated by
low-lying lines. These lines are much fainter than CO lines in the same
frequency range. The best case in point is the prominent Orion Bar PDR where
only a few low lying $\HtO$ lines are detected \citep{Habart.etal2010}, and the
$2_{1,1}-2_{0,2}$/$1_{1,0}-1_{0,1}$ luminosity ratio is 0.6; in contrast, in
$\APM{08279{+}5255}$, this ratio is $>8.1$.  Furthermore, in
$\APM{08279{+}5255}$, the $\HtO$ lines have luminosities comparable to CO lines
in the same frequency range, while they are two orders of magnitude fainter in
the Orion Bar (e.g., the $\HtO$~$2_{1,1}-2_{0,2}$~/~CO($6{-}5$) line luminosity
ratio is 0.6 in $\APM{08279{+}5255}$, while it is 0.026 in the Orion Bar).  The
only object with properties comparable to $\APM{08279{+}5255}$ is the nearby
ULIRG/QSO $\Mrk{231}$ \citep{VanDerWerf.etal2010}, where high $\HtO$ lines were
also found to be radiatively excited \citep{GonzalezAlfonso.etal2010}, and the
two objects show very similar $3_{2,1}-3_{1,2}$/$2_{0,2}-1_{1,1}$ luminosity
ratios.

The fact that in the presence of an intense far-IR radiation field the most
luminous $\HtO$ emission is radiatively excited follows directly from the large
Einstein $A$-values of the $\HtO$ rotational lines, which result in high
critical densities as well as a strong coupling to the local far-IR radiation
field.  A key result of our analysis is therefore the presence of a sufficiently
intense local far-IR radiation field, which requires the emitting clumps to be
located in an obscured environment, with $\tau_{100}\sim 0.9$.
The essential difference with the $\HtO$ excitation in the Orion Bar
PDR (aside from less important differences in temperature and density)
is the fact that the latter has a low $100\mum$ optical depth, while
in $\APM{08279{+}5255}$ $\tau_{100}\approx 0.9$ over the entire region
sampled.  The minimum size of the this region follows from the
blackbody limit, which is calculated by assigning the entire
$\LFIR=5\times10^{13}\Lsun$ (corrected for gravitational
amplification) to a blackbody with $T=220\K$. The resulting radius is
$R=110\pc$ for a spherical model or a factor $\sqrt{2}$ larger for a
disk-like configuration, as preferred by \citet{Riechers.etal2009a}.
The most recent lensing model, which accurately reproduces the
positions and luminosities of the gravitationally lensed images of the
QSO nucleus, assigns the CO(1$-$0) emission to a disk of radius of
$550\pc$, at an inclination of less than $30\deg$\ from face-on
\citep{Riechers.etal2009a}.  Because of the excellent match of the
$\HtO$ and CO line widths, it is likely that the region of $\HtO$
emission, with $\Td=220\K$ and $\tau_{100}\sim0.9$, has the same radius
as the CO emission region.

The CO rotational line ladder of $\APM{08279{+}5255}$ shows even higher
excitation than that of $\Mrk{231}$ and therefore most likely reveals the
presence of an XDR \citep{Weiss.etal2007,VanDerWerf.etal2010}. Since high $\HtO$
abundances can be generated in both PDRs and XDRs \citep{Meijerink.Spaans2005},
our data do not distinguish directly between these models. However, while X-rays
easily traverse and heat large column densities of gas, they are inefficient at
heating the dust, and dust temperatures higher than $\sim50\K$ are hard to
achieve over extended regions \citep{Yan1997,Meijerink.Spaans2005}. The high
dust temperature of $\sim220\K$ over a $\sim550\pc$ radius region is more easily
achieved in PDRs, generated by widespread star formation in clumps of dense gas
throughout the circumnuclear gas disk.  Regions of several $100\pc$ radius, with
high $100\mum$ optical depths, are locally found only in the central regions of
ULIRGs \citep{Solomon.etal1997}.  The fact that in this study we find such a
region in a QSO, provides support for the scenario where ULIRGs form the
birthplaces of QSOs \citep{Sanders.etal1988}.

If our suggestion that the extended warm dust continuum is generated
by circumnuclear star formation in $\APM{08279{+}5255}$ is correct,
this star formation is taking place in the presence of the strong and
penetrative X-ray radiation field generated by the AGN
\citep{Gallagher.etal2002,Hasinger.etal2002,Chartas.etal2002}. As
shown by \citet{Hocuk.Spaans2010}, in this situation fragmentation is
inhibited and a top-heavy stellar Initial Mass Function is expected to
result. It is possible that this effect accounts for the extraordinary
far-IR luminosity $\LFIR=5\times10^{13}\Lsun$ (corrected for
gravitational amplification) of $\APM{08279{+}5255}$. We note also
that in the clumps in our model, both turbulent pressure
$\qu{P}{turb}\sim \rho\sigma_v^2/3$ (where $\rho$ is the mass density
of the gas clump and $\sigma_v$ its turbulent velocity dispersion) and
radiation pressure $\qu{P}{rad}\sim \tau_{100}\sigma\Td^4/c$ (where
$\sigma$ is the Stefan-Boltzmann constant and $c$ the speed of light)
exceed the thermal pressure $\qu{P}{th}\sim \nHt k\Tg$ by a large
factor.  Inserting numbers from our best-fit model, we find that
$\qu{P}{th}\sim 3\times 10^{-8}\un{erg}\pcmcub$, while
$\qu{P}{turb}\sim\qu{P}{rad}\sim3\times
10^{-7}\un{erg}\pcmcub$. Radiation pressure therefore plays an
important role in the dynamics of the circumnuclear gas cloud. If the
clumps are indeed forming stars, as we are suggesting, the star
formation process is then close to Eddington-limited, in agreement
with the model developed by \citet{Thompson.etal2005} for starburst
regions surrounding a supermassive black hole.

Our study demonstrates how radiatively exited $\HtO$ lines can be used to reveal
the presence of extended infrared-opaque regions in galactic nuclei (even
without spatially resolving these regions). Furthermore, we can derive local
conditions in the infrared-opaque nuclear gas disk, which in the present case
indicates close to Eddington-limited star formation.  While for local galaxies
observations from space will remain necessary to observe the $\HtO$ lines, the
Atacama Large Millimeter Array will make this diagnostic readily available in
galaxies with sufficient redshift, without the aid of gravitational lensing.

\acknowledgements
\acknowledgeIRAMPdBI
DR acknowledges support from NASA through a Spitzer grant.
We thank Melanie Krips for expert assistance with the IRAM data reduction, and
Rodrigo Ibata for providing the NICMOS image used in \figref{fig.map}. We also
thank Xander Tielens for commenting on an earlier version of this paper.

{\it Facilities:} \facility{IRAM:Interferometer}


\begin{thebibliography}{40}
\expandafter\ifx\csname natexlab\endcsname\relax\def\natexlab#1{#1}\fi

\bibitem[{{Bradford} {et~al.}(2011){Bradford}, {Bolatto}, {Maloney}, {Aguirre},
  {Bock}, {Glenn}, {Kamenetzky}, {Lupu}, {Matsuhara}, {Murphy}, {Naylor},
  {Nguyen}, {Scott}, \& {Zmuidzinas}}]{Bradford.etal2011}
{Bradford}, C.~M., {Bolatto}, A.~D., {Maloney}, P.~R., {Aguirre}, J.~E.,
  {Bock}, J.~J., {Glenn}, J., {Kamenetzky}, J., {Lupu}, R., {Matsuhara}, H.,
  {Murphy}, E.~J., {Naylor}, B.~J., {Nguyen}, H.~T., {Scott}, K., \&
  {Zmuidzinas}, J. 2011, ArXiv e-prints

\bibitem[{{Chartas} {et~al.}(2002){Chartas}, {Brandt}, {Gallagher}, \&
  {Garmire}}]{Chartas.etal2002}
{Chartas}, G., {Brandt}, W.~N., {Gallagher}, S.~C., \& {Garmire}, G.~P. 2002,
  \apj, 579, 169

\bibitem[{{Clark}(1980)}]{Clark1980}
{Clark}, B.~G. 1980, \aap, 89, 377

\bibitem[{{Gallagher} {et~al.}(2002){Gallagher}, {Brandt}, {Chartas}, \&
  {Garmire}}]{Gallagher.etal2002}
{Gallagher}, S.~C., {Brandt}, W.~N., {Chartas}, G., \& {Garmire}, G.~P. 2002,
  \apj, 567, 37

\bibitem[{{Garc{\'{\i}}a-Burillo} {et~al.}(2006){Garc{\'{\i}}a-Burillo},
  {Graci{\'a}-Carpio}, {Gu{\'e}lin}, {Neri}, {Cox}, {Planesas}, {Solomon},
  {Tacconi}, \& {Vanden Bout}}]{GarciaBurillo.etal2006}
{Garc{\'{\i}}a-Burillo}, S., {Graci{\'a}-Carpio}, J., {Gu{\'e}lin}, M., {Neri},
  R., {Cox}, P., {Planesas}, P., {Solomon}, P.~M., {Tacconi}, L.~J., \& {Vanden
  Bout}, P.~A. 2006, \apjl, 645, L17

\bibitem[{{Gonz{\'a}lez-Alfonso} {et~al.}(2010){Gonz{\'a}lez-Alfonso},
  {Fischer}, {Isaak}, {Rykala}, {Savini}, {Spaans}, {van der Werf},
  {Meijerink}, {Israel}, {Loenen}, {Vlahakis}, {Smith}, {Charmandaris},
  {Aalto}, {Henkel}, {Wei{\ss}}, {Walter}, {Greve}, {Mart{\'{\i}}n-Pintado},
  {Naylor}, {Spinoglio}, {Veilleux}, {Harris}, {Armus}, {Lord}, {Mazzarella},
  {Xilouris}, {Sanders}, {Dasyra}, {Wiedner}, {Kramer}, {Papadopoulos},
  {Stacey}, {Evans}, \& {Gao}}]{GonzalezAlfonso.etal2010}
{Gonz{\'a}lez-Alfonso}, E., {Fischer}, J., {Isaak}, K., {Rykala}, A., {Savini},
  G., {Spaans}, M., {van der Werf}, P., {Meijerink}, R., {Israel}, F.~P.,
  {Loenen}, A.~F., {Vlahakis}, C., {Smith}, H.~A., {Charmandaris}, V., {Aalto},
  S., {Henkel}, C., {Wei{\ss}}, A., {Walter}, F., {Greve}, T.~R.,
  {Mart{\'{\i}}n-Pintado}, J., {Naylor}, D.~A., {Spinoglio}, L., {Veilleux},
  S., {Harris}, A.~I., {Armus}, L., {Lord}, S., {Mazzarella}, J., {Xilouris},
  E.~M., {Sanders}, D.~B., {Dasyra}, K.~M., {Wiedner}, M.~C., {Kramer}, C.,
  {Papadopoulos}, P.~P., {Stacey}, G.~J., {Evans}, A.~S., \& {Gao}, Y. 2010,
  \aap, 518, L43+

\bibitem[{{Guilloteau} {et~al.}(1992){Guilloteau}, {Delannoy}, {Downes},
  {Greve}, {Guelin}, {Lucas}, {Morris}, {Radford}, {Wink}, {Cernicharo},
  {Forveille}, {Garcia-Burillo}, {Neri}, {Blondel}, {Perrigourad}, {Plathner},
  \& {Torres}}]{Guilloteau.etal1992}
{Guilloteau}, S., {Delannoy}, J., {Downes}, D., {Greve}, A., {Guelin}, M.,
  {Lucas}, R., {Morris}, D., {Radford}, S.~J.~E., {Wink}, J., {Cernicharo}, J.,
  {Forveille}, T., {Garcia-Burillo}, S., {Neri}, R., {Blondel}, J.,
  {Perrigourad}, A., {Plathner}, D., \& {Torres}, M. 1992, \aap, 262, 624

\bibitem[{{Habart} {et~al.}(2010){Habart}, {Dartois}, {Abergel}, {Baluteau},
  {Naylor}, {Polehampton}, {Joblin}, {Ade}, {Anderson}, {Andr{\'e}}, {Arab},
  {Bernard}, {Blagrave}, {Bontemps}, {Boulanger}, {Cohen}, {Compiegne}, {Cox},
  {Davis}, {Emery}, {Fulton}, {Gry}, {Huang}, {Jones}, {Kirk}, {Lagache},
  {Lim}, {Madden}, {Makiwa}, {Martin}, {Miville-Desch{\^e}nes}, {Molinari},
  {Moseley}, {Motte}, {Okumura}, {Pinheiro Gon{\c c}alves}, {Rodon}, {Russeil},
  {Saraceno}, {Sidher}, {Spencer}, {Swinyard}, {Ward-Thompson}, {White}, \&
  {Zavagno}}]{Habart.etal2010}
{Habart}, E., {Dartois}, E., {Abergel}, A., {Baluteau}, J., {Naylor}, D.,
  {Polehampton}, E., {Joblin}, C., {Ade}, P., {Anderson}, L.~D., {Andr{\'e}},
  P., {Arab}, H., {Bernard}, J., {Blagrave}, K., {Bontemps}, S., {Boulanger},
  F., {Cohen}, M., {Compiegne}, M., {Cox}, P., {Davis}, G., {Emery}, R.,
  {Fulton}, T., {Gry}, C., {Huang}, M., {Jones}, S.~C., {Kirk}, J., {Lagache},
  G., {Lim}, T., {Madden}, S., {Makiwa}, G., {Martin}, P.,
  {Miville-Desch{\^e}nes}, M., {Molinari}, S., {Moseley}, H., {Motte}, F.,
  {Okumura}, K., {Pinheiro Gon{\c c}alves}, D., {Rodon}, J., {Russeil}, D.,
  {Saraceno}, P., {Sidher}, S., {Spencer}, L., {Swinyard}, B., {Ward-Thompson},
  D., {White}, G.~J., \& {Zavagno}, A. 2010, \aap, 518, L116+

\bibitem[{{Hasinger} {et~al.}(2002){Hasinger}, {Schartel}, \&
  {Komossa}}]{Hasinger.etal2002}
{Hasinger}, G., {Schartel}, N., \& {Komossa}, S. 2002, \apjl, 573, L77

\bibitem[{{Hocuk} \& {Spaans}(2010)}]{Hocuk.Spaans2010}
{Hocuk}, S. \& {Spaans}, M. 2010, \aap, 522, A24+

\bibitem[{{Hollenbach} {et~al.}(2009){Hollenbach}, {Kaufman}, {Bergin}, \&
  {Melnick}}]{Hollenbach.etal2009}
{Hollenbach}, D., {Kaufman}, M.~J., {Bergin}, E.~A., \& {Melnick}, G.~J. 2009,
  \apj, 690, 1497

\bibitem[{{Ibata} {et~al.}(1999){Ibata}, {Lewis}, {Irwin}, {Leh{\'a}r}, \&
  {Totten}}]{Ibata.etal1999}
{Ibata}, R.~A., {Lewis}, G.~F., {Irwin}, M.~J., {Leh{\'a}r}, J., \& {Totten},
  E.~J. 1999, \aj, 118, 1922

\bibitem[{{Impellizzeri} {et~al.}(2008){Impellizzeri}, {McKean}, {Castangia},
  {Roy}, {Henkel}, {Brunthaler}, \& {Wucknitz}}]{Impellizzeri.etal2008}
{Impellizzeri}, C.~M.~V., {McKean}, J.~P., {Castangia}, P., {Roy}, A.~L.,
  {Henkel}, C., {Brunthaler}, A., \& {Wucknitz}, O. 2008, \nat, 456, 927

\bibitem[{{Irwin} {et~al.}(1998){Irwin}, {Ibata}, {Lewis}, \&
  {Totten}}]{Irwin.etal1998}
{Irwin}, M.~J., {Ibata}, R.~A., {Lewis}, G.~F., \& {Totten}, E.~J. 1998, \apj,
  505, 529

\bibitem[{{Krips} {et~al.}(2007){Krips}, {Peck}, {Sakamoto}, {Petitpas},
  {Wilner}, {Matsushita}, \& {Iono}}]{Krips.etal2007}
{Krips}, M., {Peck}, A.~B., {Sakamoto}, K., {Petitpas}, G.~B., {Wilner}, D.~J.,
  {Matsushita}, S., \& {Iono}, D. 2007, \apjl, 671, L5

\bibitem[{{Lewis} {et~al.}(1998){Lewis}, {Chapman}, {Ibata}, {Irwin}, \&
  {Totten}}]{Lewis.etal1998}
{Lewis}, G.~F., {Chapman}, S.~C., {Ibata}, R.~A., {Irwin}, M.~J., \& {Totten},
  E.~J. 1998, \apjl, 505, L1+

\bibitem[{{Lis} {et~al.}(2011){Lis}, {Neufeld}, {Phillips}, {Gerin}, \&
  {Neri}}]{Lis.etal2011}
{Lis}, D.~C., {Neufeld}, D.~A., {Phillips}, T.~G., {Gerin}, M., \& {Neri}, R.
  2011, ArXiv e-prints

\bibitem[{{Meijerink} \& {Spaans}(2005)}]{Meijerink.Spaans2005}
{Meijerink}, R. \& {Spaans}, M. 2005, \aap, 436, 397

\bibitem[{{Neufeld} \& {Kaufman}(1993)}]{Neufeld.Kaufman1993}
{Neufeld}, D.~A. \& {Kaufman}, M.~J. 1993, \apj, 418, 263

\bibitem[{{Neufeld} {et~al.}(1995){Neufeld}, {Lepp}, \&
  {Melnick}}]{Neufeld.etal1995}
{Neufeld}, D.~A., {Lepp}, S., \& {Melnick}, G.~J. 1995, \apjs, 100, 132

\bibitem[{{Omont} {et~al.}(2011){Omont}, {Neri}, {Cox}, {Lupu}, {Gu{\'e}lin},
  {van der Werf}, {Wei{\ss}}, {Ivison}, {Negrello}, {Leeuw}, {Lehnert},
  {Smail}, {Verma}, {Baker}, {Beelen}, {Aguirre}, {Baes}, {Bertoldi},
  {Clements}, {Cooray}, {Coppin}, {Dannerbauer}, {de Zotti}, {Dye}, {Fiolet},
  {Frayer}, {Gavazzi}, {Hughes}, {Jarvis}, {Krips}, {Micha{\l}owski}, {Murphy},
  {Riechers}, {Serjeant}, {Swinbank}, {Temi}, {Vaccari}, {Vieira}, {Auld},
  {Buttiglione}, {Cava}, {Dariush}, {Dunne}, {Eales}, {Fritz}, {Gomez}, {Ibar},
  {Maddox}, {Pascale}, {Pohlen}, {Rigby}, {Smith}, {Bock}, {Bradford}, {Glenn},
  {Scott}, \& {Zmuidzinas}}]{Omont.etal2011}
{Omont}, A., {Neri}, R., {Cox}, P., {Lupu}, R., {Gu{\'e}lin}, M., {van der
  Werf}, P., {Wei{\ss}}, A., {Ivison}, R., {Negrello}, M., {Leeuw}, L.,
  {Lehnert}, M., {Smail}, I., {Verma}, A., {Baker}, A.~J., {Beelen}, A.,
  {Aguirre}, J.~E., {Baes}, M., {Bertoldi}, F., {Clements}, D.~L., {Cooray},
  A., {Coppin}, K., {Dannerbauer}, H., {de Zotti}, G., {Dye}, S., {Fiolet}, N.,
  {Frayer}, D., {Gavazzi}, R., {Hughes}, D., {Jarvis}, M., {Krips}, M.,
  {Micha{\l}owski}, M.~J., {Murphy}, E.~J., {Riechers}, D., {Serjeant}, S.,
  {Swinbank}, A.~M., {Temi}, P., {Vaccari}, M., {Vieira}, J.~D., {Auld}, R.,
  {Buttiglione}, B., {Cava}, A., {Dariush}, A., {Dunne}, L., {Eales}, S.~A.,
  {Fritz}, J., {Gomez}, H., {Ibar}, E., {Maddox}, S., {Pascale}, E., {Pohlen},
  M., {Rigby}, E., {Smith}, D.~J.~B., {Bock}, J., {Bradford}, C.~M., {Glenn},
  J., {Scott}, K.~S., \& {Zmuidzinas}, J. 2011, \aap, 530, L3+

\bibitem[{{Panuzzo} {et~al.}(2010){Panuzzo}, {Rangwala}, {Rykala}, {Isaak},
  {Glenn}, {Wilson}, {Auld}, {Baes}, {Barlow}, {Bendo}, {Bock}, {Boselli},
  {Bradford}, {Buat}, {Castro-Rodr{\'{\i}}guez}, {Chanial}, {Charlot},
  {Ciesla}, {Clements}, {Cooray}, {Cormier}, {Cortese}, {Davies}, {Dwek},
  {Eales}, {Elbaz}, {Fulton}, {Galametz}, {Galliano}, {Gear}, {Gomez},
  {Griffin}, {Hony}, {Levenson}, {Lu}, {Madden}, {O'Halloran}, {Okumura},
  {Oliver}, {Page}, {Papageorgiou}, {Parkin}, {P{\'e}rez-Fournon}, {Pohlen},
  {Polehampton}, {Rigby}, {Roussel}, {Sacchi}, {Sauvage}, {Schulz}, {Schirm},
  {Smith}, {Spinoglio}, {Stevens}, {Srinivasan}, {Symeonidis}, {Swinyard},
  {Trichas}, {Vaccari}, {Vigroux}, {Wozniak}, {Wright}, \&
  {Zeilinger}}]{Panuzzo.etal2010}
{Panuzzo}, P., {Rangwala}, N., {Rykala}, A., {Isaak}, K.~G., {Glenn}, J.,
  {Wilson}, C.~D., {Auld}, R., {Baes}, M., {Barlow}, M.~J., {Bendo}, G.~J.,
  {Bock}, J.~J., {Boselli}, A., {Bradford}, M., {Buat}, V.,
  {Castro-Rodr{\'{\i}}guez}, N., {Chanial}, P., {Charlot}, S., {Ciesla}, L.,
  {Clements}, D.~L., {Cooray}, A., {Cormier}, D., {Cortese}, L., {Davies},
  J.~I., {Dwek}, E., {Eales}, S.~A., {Elbaz}, D., {Fulton}, T., {Galametz}, M.,
  {Galliano}, F., {Gear}, W.~K., {Gomez}, H.~L., {Griffin}, M., {Hony}, S.,
  {Levenson}, L.~R., {Lu}, N., {Madden}, S., {O'Halloran}, B., {Okumura}, K.,
  {Oliver}, S., {Page}, M.~J., {Papageorgiou}, A., {Parkin}, T.~J.,
  {P{\'e}rez-Fournon}, I., {Pohlen}, M., {Polehampton}, E.~T., {Rigby}, E.~E.,
  {Roussel}, H., {Sacchi}, N., {Sauvage}, M., {Schulz}, B., {Schirm}, M.~R.~P.,
  {Smith}, M.~W.~L., {Spinoglio}, L., {Stevens}, J.~A., {Srinivasan}, S.,
  {Symeonidis}, M., {Swinyard}, B., {Trichas}, M., {Vaccari}, M., {Vigroux},
  L., {Wozniak}, H., {Wright}, G.~S., \& {Zeilinger}, W.~W. 2010, \aap, 518,
  L37+

\bibitem[{{Poelman} \& {Spaans}(2005)}]{Poelman.Spaans2005}
{Poelman}, D.~R. \& {Spaans}, M. 2005, \aap, 440, 559

\bibitem[{{Poelman} \& {Spaans}(2006)}]{Poelman.Spaans2006}
---. 2006, \aap, 453, 615

\bibitem[{{Riechers} {et~al.}(2009{\natexlab{a}}){Riechers}, {Walter},
  {Bertoldi}, {Carilli}, {Aravena}, {Neri}, {Cox}, {Wei{\ss}}, \&
  {Menten}}]{Riechers.etal2009b}
{Riechers}, D.~A., {Walter}, F., {Bertoldi}, F., {Carilli}, C.~L., {Aravena},
  M., {Neri}, R., {Cox}, P., {Wei{\ss}}, A., \& {Menten}, K.~M.
  2009{\natexlab{a}}, \apj, 703, 1338

\bibitem[{{Riechers} {et~al.}(2009{\natexlab{b}}){Riechers}, {Walter},
  {Carilli}, \& {Lewis}}]{Riechers.etal2009a}
{Riechers}, D.~A., {Walter}, F., {Carilli}, C.~L., \& {Lewis}, G.~F.
  2009{\natexlab{b}}, \apj, 690, 463

\bibitem[{{Riechers} {et~al.}(2006){Riechers}, {Weiss}, {Walter}, {Carilli}, \&
  {Knudsen}}]{Riechers.etal2006}
{Riechers}, D.~A., {Weiss}, A., {Walter}, F., {Carilli}, C.~L., \& {Knudsen},
  K.~K. 2006, \apj, 649, 635

\bibitem[{{Riechers} {et~al.}(2010){Riechers}, {Wei{\ss}}, {Walter}, \&
  {Wagg}}]{Riechers.etal2010}
{Riechers}, D.~A., {Wei{\ss}}, A., {Walter}, F., \& {Wagg}, J. 2010, \apj, 725,
  1032

\bibitem[{{Sanders} {et~al.}(1988){Sanders}, {Soifer}, {Elias}, {Madore},
  {Matthews}, {Neugebauer}, \& {Scoville}}]{Sanders.etal1988}
{Sanders}, D.~B., {Soifer}, B.~T., {Elias}, J.~H., {Madore}, B.~F., {Matthews},
  K., {Neugebauer}, G., \& {Scoville}, N.~Z. 1988, \apj, 325, 74

\bibitem[{{Solomon} {et~al.}(1997){Solomon}, {Downes}, {Radford}, \&
  {Barrett}}]{Solomon.etal1997}
{Solomon}, P.~M., {Downes}, D., {Radford}, S.~J.~E., \& {Barrett}, J.~W. 1997,
  \apj, 478, 144

\bibitem[{{Srianand} \& {Petitjean}(2000)}]{Srianand.Petitjean2000}
{Srianand}, R. \& {Petitjean}, P. 2000, \aap, 357, 414

\bibitem[{{Thompson} {et~al.}(2005){Thompson}, {Quataert}, \&
  {Murray}}]{Thompson.etal2005}
{Thompson}, T.~A., {Quataert}, E., \& {Murray}, N. 2005, \apj, 630, 167

\bibitem[{{Tielens} {et~al.}(1991){Tielens}, {Tokunaga}, {Geballe}, \&
  {Baas}}]{Tielens.etal1991}
{Tielens}, A.~G.~G.~M., {Tokunaga}, A.~T., {Geballe}, T.~R., \& {Baas}, F.
  1991, \apj, 381, 181

\bibitem[{{Van der Werf} {et~al.}(2010){Van der Werf}, {Isaak}, {Meijerink},
  {Spaans}, {Rykala}, {Fulton}, {Loenen}, {Walter}, {Wei{\ss}}, {Armus},
  {Fischer}, {Israel}, {Harris}, {Veilleux}, {Henkel}, {Savini}, {Lord},
  {Smith}, {Gonz{\'a}lez-Alfonso}, {Naylor}, {Aalto}, {Charmandaris}, {Dasyra},
  {Evans}, {Gao}, {Greve}, {G{\"u}sten}, {Kramer}, {Mart{\'{\i}}n-Pintado},
  {Mazzarella}, {Papadopoulos}, {Sanders}, {Spinoglio}, {Stacey}, {Vlahakis},
  {Wiedner}, \& {Xilouris}}]{VanDerWerf.etal2010}
{Van der Werf}, P.~P., {Isaak}, K.~G., {Meijerink}, R., {Spaans}, M., {Rykala},
  A., {Fulton}, T., {Loenen}, A.~F., {Walter}, F., {Wei{\ss}}, A., {Armus}, L.,
  {Fischer}, J., {Israel}, F.~P., {Harris}, A.~I., {Veilleux}, S., {Henkel},
  C., {Savini}, G., {Lord}, S., {Smith}, H.~A., {Gonz{\'a}lez-Alfonso}, E.,
  {Naylor}, D., {Aalto}, S., {Charmandaris}, V., {Dasyra}, K.~M., {Evans}, A.,
  {Gao}, Y., {Greve}, T.~R., {G{\"u}sten}, R., {Kramer}, C.,
  {Mart{\'{\i}}n-Pintado}, J., {Mazzarella}, J., {Papadopoulos}, P.~P.,
  {Sanders}, D.~B., {Spinoglio}, L., {Stacey}, G., {Vlahakis}, C., {Wiedner},
  M.~C., \& {Xilouris}, E.~M. 2010, \aap, 518, L42+

\bibitem[{{Wagg} {et~al.}(2005){Wagg}, {Wilner}, {Neri}, {Downes}, \&
  {Wiklind}}]{Wagg.etal2005}
{Wagg}, J., {Wilner}, D.~J., {Neri}, R., {Downes}, D., \& {Wiklind}, T. 2005,
  \apjl, 634, L13

\bibitem[{{Wagg} {et~al.}(2006){Wagg}, {Wilner}, {Neri}, {Downes}, \&
  {Wiklind}}]{Wagg.etal2006}
---. 2006, \apj, 651, 46

\bibitem[{{Wei{\ss}} {et~al.}(2007){Wei{\ss}}, {Downes}, {Neri}, {Walter},
  {Henkel}, {Wilner}, {Wagg}, \& {Wiklind}}]{Weiss.etal2007}
{Wei{\ss}}, A., {Downes}, D., {Neri}, R., {Walter}, F., {Henkel}, C., {Wilner},
  D.~J., {Wagg}, J., \& {Wiklind}, T. 2007, \aap, 467, 955

\bibitem[{{Wei{\ss}} {et~al.}(2010){Wei{\ss}}, {Requena-Torres}, {G{\"u}sten},
  {Garc{\'{\i}}a-Burillo}, {Harris}, {Israel}, {Klein}, {Kramer}, {Lord},
  {Martin-Pintado}, {R{\"o}llig}, {Stutzki}, {Szczerba}, {van der Werf},
  {Philipp-May}, {Yorke}, {Akyilmaz}, {Gal}, {Higgins}, {Marston}, {Roberts},
  {Schl{\"o}der}, {Schultz}, {Teyssier}, {Whyborn}, \&
  {Wunsch}}]{Weiss.etal2010}
{Wei{\ss}}, A., {Requena-Torres}, M.~A., {G{\"u}sten}, R.,
  {Garc{\'{\i}}a-Burillo}, S., {Harris}, A.~I., {Israel}, F.~P., {Klein}, T.,
  {Kramer}, C., {Lord}, S., {Martin-Pintado}, J., {R{\"o}llig}, M., {Stutzki},
  J., {Szczerba}, R., {van der Werf}, P.~P., {Philipp-May}, S., {Yorke}, H.,
  {Akyilmaz}, M., {Gal}, C., {Higgins}, R., {Marston}, A., {Roberts}, J.,
  {Schl{\"o}der}, F., {Schultz}, M., {Teyssier}, D., {Whyborn}, N., \&
  {Wunsch}, H.~J. 2010, \aap, 521, L1+

\bibitem[{{White} {et~al.}(2010){White}, {Abergel}, {Spencer}, {Schneider},
  {Naylor}, {Anderson}, {Joblin}, {Ade}, {Andr{\'e}}, {Arab}, {Baluteau},
  {Bernard}, {Blagrave}, {Bontemps}, {Boulanger}, {Cohen}, {Compiegne}, {Cox},
  {Dartois}, {Davis}, {Emery}, {Fulton}, {Gom}, {Griffin}, {Gry}, {Habart},
  {Huang}, {Jones}, {Kirk}, {Lagache}, {Leeks}, {Lim}, {Madden}, {Makiwa},
  {Martin}, {Miville-Desch{\^e}nes}, {Molinari}, {Moseley}, {Motte}, {Okumura},
  {Pinheiro Gon{\c c}alves}, {Polehampton}, {Rodet}, {Rod{\'o}n}, {Russeil},
  {Saraceno}, {Sidher}, {Swinyard}, {Ward-Thompson}, \&
  {Zavagno}}]{White.etal2010}
{White}, G.~J., {Abergel}, A., {Spencer}, L., {Schneider}, N., {Naylor}, D.~A.,
  {Anderson}, L.~D., {Joblin}, C., {Ade}, P., {Andr{\'e}}, P., {Arab}, H.,
  {Baluteau}, J.-P., {Bernard}, J.-P., {Blagrave}, K., {Bontemps}, S.,
  {Boulanger}, F., {Cohen}, M., {Compiegne}, M., {Cox}, P., {Dartois}, E.,
  {Davis}, G., {Emery}, R., {Fulton}, T., {Gom}, B., {Griffin}, M., {Gry}, C.,
  {Habart}, E., {Huang}, M., {Jones}, S., {Kirk}, J.~M., {Lagache}, G.,
  {Leeks}, S., {Lim}, T., {Madden}, S., {Makiwa}, G., {Martin}, P.,
  {Miville-Desch{\^e}nes}, M.-A., {Molinari}, S., {Moseley}, H., {Motte}, F.,
  {Okumura}, K., {Pinheiro Gon{\c c}alves}, D., {Polehampton}, E., {Rodet}, T.,
  {Rod{\'o}n}, J.~A., {Russeil}, D., {Saraceno}, P., {Sidher}, S., {Swinyard},
  B.~M., {Ward-Thompson}, D., \& {Zavagno}, A. 2010, \aap, 518, L114+

\bibitem[{{Yan}(1997)}]{Yan1997}
{Yan}, M. 1997, PhD thesis, Harvard University

\end{thebibliography}
\end{document}